\documentclass{mem}
\usepackage{natbib}\usepackage{txfonts}\usepackage{balance}
\usepackage{graphicx}
\usepackage[a4paper]{hyperref}
\idline{1}{1}
\begin{document}
\def\teff{$T\rm_{eff }$}
\def\kms{$\mathrm {km s}^{-1}$}

\title{
Cepheid distances from interferometry
}

   \subtitle{}

\author{
P. Kervella\inst{1} 
          }

\offprints{P. Kervella}
 
\institute{
LESIA, UMR\,8109, Observatoire de Paris-Meudon, 5, place Jules Janssen, F-92195 Meudon Cedex, France.
\email{pierre.kervella@obspm.fr}
}

\authorrunning{Kervella}

\titlerunning{Cepheid distances from interferometry}

\abstract{
Long baseline interferometry is now able to resolve the pulsational change of the angular diameter of a significant number of Cepheids in the solar neighborhood. This allows the application of a new version of the Baade-Wesselink (BW) method to measure their distance, for which we do not need to estimate the star's temperature. Using angular diameter measurements from the VLT Interferometer, we derived the distances to four nearby Cepheids. For three additional stars, we obtained average values of their angular diameters. Based on these new measurements and already existing data, we derived calibrations of the Period-Luminosity and Period-Radius relations. We also obtained reliable surface brightness-color relations, that can be employed for the infrared surface brightness version of the BW method.
\keywords{Stars: variables: Cepheids, Techniques: interferometric, Stars: oscillations}
}
\maketitle{}

\section{The interferometric Baade-Wesselink method}

The basic principle of the BW method is to compare the linear and angular size variation of a pulsating star, in order to derive its distance through a simple division. This is a well-established way to determine the luminosity and radius of a pulsating star. The two quantities required are the radius variation curve, that is integrated from the radial velocity curve, and the angular diameter variation.

On one hand, the linear size variation can be obtained by high resolution spectroscopy, through the integration of the radial velocity curve obtained by monitoring the Doppler shift of the spectral lines present in the spectrum. A difficulty in this process is that the measured wavelength shifts are integrated values over the full stellar disk.  To convert them into a pulsation velocity, i.e. a physical displacement of the photosphere at the center of the disk, we have to multiply it by a projection factor $p$ that encompasses the sphericity of the star and the structure of its atmosphere (limb darkening,É). The $p$-factor is still uncertain at a level of a few percents, therefore limiting the Period-Luminosity (P--L) calibration accuracy at this level. The situation has improved recently with the work of M\'erand et al.~\cite{merand05}, reported in this volume, who measured the $p$-factor of the nearby Cepheid $\delta$\,Cep by interferometry, taking advantage of the availability of a precise trigonometric parallax (Benedict et al.~\cite{benedict02}).

On the other hand, the angular size is difficult to estimate directly. Until recently, the only method available was to estimate the surface brightness of the star. With the advent of infrared long baseline interferometers, it is now possible to resolve spatially the star itself, and thus measure directly its photospheric angular diameter. An uncertainty at a level of about 1\% remains on the limb darkening of these stars, that is currently taken from static atmosphere models.

\section{Observations with the VLTI}

For our observations, the beams from the two VLTI Test Siderostats (0.35\,m aperture) or the two Unit Telescopes UT1 and UT3 were recombined coherently in VINCI, the VLTI Commissioning Instrument. We used a regular $K$ band filter ($\lambda = 2.0-2.4\,\mu$m) that gives an effective observation wavelength of 2.18\,$\mu$m for the effective temperature of typical Cepheids. Three VLTI baselines were used for this program: E0-G1, B3-M0 and UT1-UT3 respectively 66, 140 and 102.5\,m in ground length.  In total, we obtained 69 individual angular diameter measurements, for a total of more than 100\,hours of telescope time (2\,hours with the UTs), spread over 68 nights (Kervella et al.~\cite{kervella04a}).

\begin{figure}[t!]
\resizebox{\hsize}{!}{\includegraphics[clip=true]{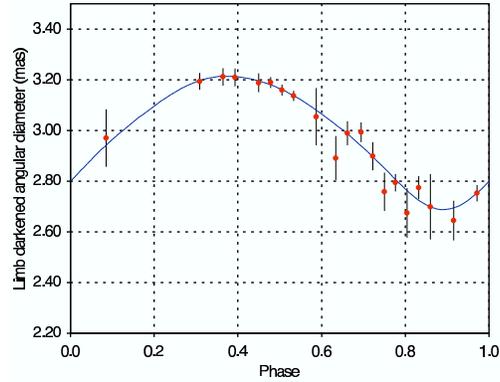}}
\caption{\footnotesize
Angular diameter variation of $\ell$\,Car measured with the VINCI/VLTI instrument.
}
\label{l_car}
\end{figure}

Considering the constraints  in terms of sky coverage, limiting magnitude and accessible resolution, we selected seven bright Cepheids observable from Paranal Observatory (latitude -24$^\circ$): X\,Sgr, $\eta$\,Aql, W\,Sgr, $\beta$\,Dor, $\zeta$\,Gem, Y\,Oph and $\ell$\,Car. The periods of these stars cover a wide range, from 7 to 35.5\,days, an important advantage to properly constrain the P--R and P--L relations. Using the interferometric BW method, we derived the distances to $\eta$\,Aql, W\,Sgr, $\beta$\,Dor and $\ell$\,Car. For the remaining three objects of our sample, X\,Sgr, $\zeta$\,Gem and Y\,Oph, we obtained average values of their angular diameters, and we applied a hybrid method to derive their distances, based on published values of their linear diameters. Fig.~\ref{l_car} shows the angular diameter curve and the fitted radius curve of $\ell$\,Car ($P = 35.5$\,days), that constrains its distance to a relative precision better than 5\%. A discussion of these data can be found in Kervella et al.~\cite{kervella04d}.

\section{Period-Radius relation}

The P--R relation is an important constraint to the Cepheid models (see e.g. Bono et al.~\cite{bono98}). It takes the form of the linear expression $\log R = a \log P + b$. In order to calibrate this relation, we need to estimate directly the linear radii of a set of Cepheids.  To complement the VINCI sample of seven Cepheids, we added the measurements of $\delta$\,Cep, $\zeta$\,Gem and $\eta$\,Aql obtained previously by other interferometers. We have applied two methods to determine the radii of the Cepheids of our sample: the interferometric BW method, and a combination of the average angular diameter and trigonometric parallax. While the first provides directly the average linear radius and distance, we need to use trigonometric parallaxes to derive the radii of the Cepheids for which the pulsation is not detected.  For these stars, we applied the Hipparcos distance, except for $\delta$\,Cep, for which we considered the recent parallax measurement by Benedict et al.~\cite{benedict02}.

\begin{figure}[t!]
\resizebox{\hsize}{!}{\includegraphics[clip=true]{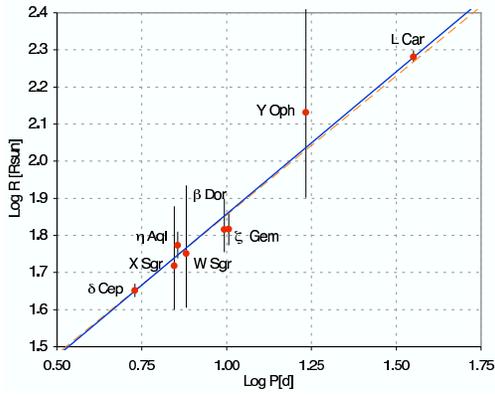}}
\caption{\footnotesize
Period-Radius relation deduced from interferometric measurements of Cepheids. The solid line results from the simultaneous fit of the zero point and slope of the line. The dashed line assumes the slope from GFG98, fitting only the zero point.
}
\label{period_radius}
\end{figure}

Fig.~\ref{period_radius} shows the distribution of the measured diameters on the P-R diagram. When we choose to consider a constant slope of $a=0.750 \pm 0.024$, as found by Gieren, Fouqu\'e \& G\`omez~\cite{gieren98}, hereafter GFG98, we derive a zero point of $b=1.105 \pm 0.029$ (Kervella et al.~\cite{kervella04b}). As a comparison, GFG98 obtained a value of $b=1.075 \pm 0.007$, only $-1.6 \sigma$ away from our result.  These relations are compatible with our calibration within their error bars. Fitting simultaneously both the slope and the zero point to our data set, we obtain $a=0.767 \pm 0.009$ and $b = 1.091 \pm 0.011$.  These values are only $\Delta a = +0.7 \sigma$ and $\Delta b= +1.2 \sigma$ away from the GFG98 calibration.  Considering the limited size of our sample, the agreement is very satisfactory.  On the other hand, the slopes derived from numerical models of Cepheids are significantly different, such as the slope $a=0.661 \pm 0.006$ found by Bono et al.~\cite{bono98}.

\section{Period-Luminosity relations}

Our sample is currently too limited to allow a robust determination of the P--L relation, defined as $M_\lambda= \alpha_\lambda (\log P - 1) + \beta_\lambda$ that would include both the slope $\alpha_\lambda$ and the $\log P =1$ reference point $\beta_\lambda$. However, if we suppose that the slope is known a priori from the literature, we can still derive a precise calibration. We have considered for our fit the P--L slope measured on LMC Cepheids (GFG98). This is a reasonable assumption, as it was checked successfully on the Magellanic Clouds Cepheids.

\begin{figure}[t!]
\resizebox{\hsize}{!}{\includegraphics[clip=true]{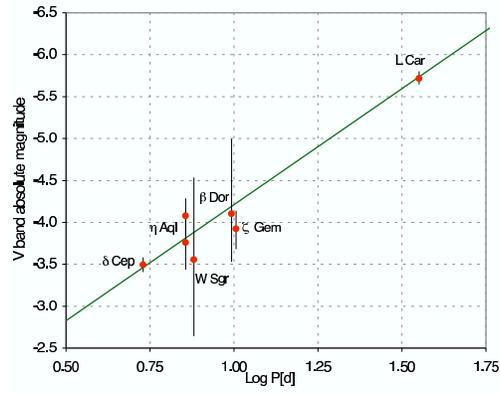}}
\caption{\footnotesize
Period-Luminosity relation in the $V$ band, as deduced from the interferometric observations of Cepheids and the HST parallax measurement of $\delta$\,Cep. The solid line is the fitted P--L relation, assuming the slope from GFG98.}
\label{period_luminosity}
\end{figure}

For the $V$ band, we obtain $\beta_V= -4.209 \pm 0.075$ (Kervella et al.~\cite{kervella04b}). The positions of the Cepheids on the P--L diagram are shown on Fig.~Ê\ref{period_luminosity}. Our calibrations differ from GFG98 by $\Delta \beta_V = +0.14$\,mag, corresponding to $+1.8\sigma$, respectively. The sample is dominated by the high precision $\ell$\,Car and $\delta$\,Cep measurements.
The two stars $\ell$\,Car and $\delta$\,Cep do not appear to be systematically different from the other Cepheids of our sample. It is difficult to conclude firmly to a significant discrepancy between GFG98 and our results, as our sample is currently too limited to exclude a small-statistics bias. However, if we assume an intrinsic dispersion of the P--L relation $\sigma=0.1$\,mag, as suggested by GFG98, then our results point toward a slight underestimation of the absolute magnitudes of Cepheids by these authors.  On the other hand, we obtain precisely the same $\log P = 1$ reference point value in $V$ as Lanoix et al.~\cite{lanoix99} did, using Hipparcos parallaxes.  The excellent agreement between these two independent calibrations is quite remarkable.

\section{Surface brightness relations}

The surface brightness-color relations (hereafter SB, see e.g. Fouqu\'e, Storm \& Gieren~\cite{fouque03}) are of considerable astrophysical interest for Cepheids, as a well-defined relation between a color index and the surface brightness can provide accurate predictions of their angular diameters. When combined with the radius curve, integrated from spectroscopic radial velocity measurements, they give access to the distance of the Cepheid through the classical (non interferometric) BW method. This method has been applied recently to Cepheids in the SMC (Storm et al.~\cite{storm04}), i.e. at a far greater distance than what can be achieved by the interferometric BW method. But the accuracy that can be achieved on the distance estimate is conditioned for a large part by our knowledge of the SB relations.

With the hypothesis of a perfect blackbody, any color can in principle be used to obtain the SB, but in practice, the linearity of the correspondance between log Teff and color depends on the chosen wavelength bands. The surface brightness $F_\lambda$ is given by the following expression (Fouqu\'e \& Gieren~\cite{fouque97}): $F_\lambda = 4.2207 - 0.1\ m_\lambda - 0.5 \log \theta_{\rm LD}$ where $\theta_{\rm LD}$ is the limb darkened angular diameter, i.e. the angular size of the stellar photosphere. 

\begin{figure}[t!]
\resizebox{\hsize}{!}{\includegraphics[clip=true]{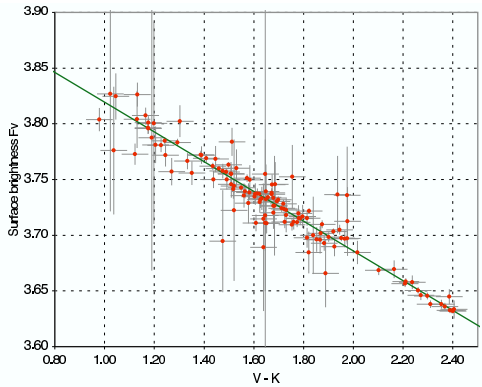}}
\caption{\footnotesize
Interferometric calibration of the $F_V(V-K)$ surface brightness-color relation of Cepheids.}
\label{surface_brightness}
\end{figure}

As an example, the resulting $F_V(V-K)$ relation fit is presented in Fig.~\ref{surface_brightness}. The smallest residual dispersions are obtained for the infrared based colors, for instance:
$F_V = -0.1336 \pm 0.0008\ (V-K) + 3.9530 \pm 0.0006$
(Kervella et al.~\cite{kervella04c}). The intrinsic dispersion is undetectable at the current level of precision of our measurements, and could be as low as 1\%.

\section{Prospects}

The field of Cepheid studies by interferometry is currently in a very dynamical phase. The availability of new, more powerful instruments in the recent years makes it possible to study in details the changing disk of these stars over their pulsation. The results presented in this paper represent a first step towards an accurate calibration of the fundamental relations of Cepheids, and the upcoming availability of 1.8\,m Auxiliary Telescopes on the VLTI platform will allow us to observe many more stars with a quality at least as good as our observations of $\ell$\,Car. Together with the spectrographic capabilities of the AMBER instrument, we will be able to resolve spatially the internal behavior of the Cepheid atmospheres. In the northern hemisphere, the CHARA Array (see M\'erand et al. in these proceedings) extends the interferometric coverage to the whole sky, with in particular an excellent view on Polaris.
 
\begin{acknowledgements}
Based on observations collected at the European Southern Observatory,
Cerro Paranal, Chile, in the framework of ESO programme 071.D-0425.
\end{acknowledgements}

\bibliographystyle{aa}

\end{document}